\documentclass[12pt,preprint]{aastex}
\newcommand{\HI}{H~{\sc i}} 

\newcommand{\kms}{${\rm km~s^{-1}}$}
\newcommand{\ha}{H${\alpha}$}

\slugcomment{To appear in {\em The Astrophysical Journal}}
\shortauthors{McCLURE-GRIFFITHS ET AL.} 
\shorttitle{}

\begin{document} 

\title{Measurement of a Magnetic Field in a Leading Arm High Velocity Cloud}

\author{N.\ M.\ McClure-Griffiths,\altaffilmark{1,2} G.\ J.\ Madsen,\altaffilmark{2} B.\ M.\ Gaensler,\altaffilmark{2} D.\ McConnell,\altaffilmark{1} \& D.\ H.\ F.\ M.\ Schnitzeler\altaffilmark{1}}

\altaffiltext{1}{Australia Telescope National Facility, CSIRO
  Astronomy \& Space Science, Marsfield NSW 2122, Australia;
  naomi.mcclure-griffiths@csiro.au, david.mcconnell@csiro.au, dominic.schnitzeler@csiro.au}
\altaffiltext{2}{Sydney Institute for Astronomy, School of Physics, The University of Sydney, NSW 2006, Australia; greg.madsen@sydney.edu.au, bryan.gaensler@sydney.edu.au}
%--------------------------------------------
\begin{abstract}
%-------------------------------------------
Using a recent catalogue of extragalactic Faraday rotation derived from the NRAO VLA Sky Survey we have found an agreement between Faraday rotation structure and the \HI\ emission structure of a High Velocity Cloud (HVC) associated with the Leading Arm of the Magellanic System.  We suggest that this morphological agreement is indicative of Faraday rotation through the HVC.  Under this assumption we have used 48 rotation measures through the HVC, together with estimates of the electron column density from \ha\ measurements and QSO absorption lines to estimate a strength for the line-of-sight component of the coherent magnetic field in the HVC of $\langle B_{\parallel}\rangle \gtrsim 6~{\rm \mu G}$.  A coherent magnetic field of this strength is more than sufficient to dynamically stabilize the cloud against ram pressure stripping by the Milky Way halo and may also provide thermal insulation for the cold cloud.   We estimate an upper limit to the ratio of random to coherent magnetic field of $B_{r}/B_{\parallel} < 0.8$, which suggests that the random field does not dominate over the coherent field as it does in the Magellanic Clouds from which this HVC likely originates.
  \end{abstract}

%----------------------------------------------------------
\section{Introduction}
\label{sec:intro}
%----------------------------------------------------------
Galaxy disks like that of the Milky Way require fuel to continue their star formation.  High velocity clouds (HVCs), first identified in 21 cm emission at anomalous (non-Galactic) velocities, have been suggested as a source of fuel \citep[e.g.][]{quilis01}.  This raises questions about how HVCs can survive their passage through the halo of the Milky Way long enough to deposit their gas onto the Galactic disk.  Many HVCs exhibit head-tail structure \citep[e.g.][]{bruens00}, showing that they are interacting with an external medium.  Such interactions could produce Rayleigh-Taylor and Kelvin-Helmholtz instabilities that would destroy the cloud in significantly less than a free-fall time \citep{benjamin97}.   Furthermore, HVCs bathed in the hot halo of the Milky Way should evaporate on a similar timescale.  It is therefore important to understand how HVCs might avoid fragmentation and evaporation long enough to deposit cold gas onto the Milky Way disk.  The surface tension associated with magnetic fields may provide the means to stabilize clouds \citep{jones96}.   Magnetohydrodynamic simulations by \cite{konz02} have shown that even small ($\sim 0.3~{\rm \mu G}$) pre-existing halo magnetic fields are compressed along the head of a moving HVC by a factor of ten or more and can provide the stability and thermal insulation required to maintain a cohesive HVC for more than a free-fall time.

However, there are no clear measurements of magnetic field strengths in HVCs.   Four HVCs have been observed for magnetic fields using Zeeman splitting of the \HI\ line \citep{kazes91}. The most promising possibility, HVC 132$+$23$+$212, suggested a field of 11 $\mu$G, but has never been confirmed.  

Magnetic fields in HVCs may also be evident in Faraday rotation measurements towards background extragalactic polarized sources.  Faraday rotation of the polarized synchrotron radiation from radio sources probes the product of the magnetic field, $B_{\parallel}$ (${\rm \mu G}$), and electron density, $n_e(l)$ (${\rm cm^{-3}}$), along the line of sight, $l$ (pc), such that the measured rotation in linear polarization angle is $\Delta \theta = RM \lambda^2$, where the rotation measure, $RM$, is:
\begin{equation}
RM = 0.81 \int_{source}^{observer} n_e(l) \, B_{\parallel} \, dl \,\, {\rm rad~m^{-2 }}.
\label{eq:rm}
\end{equation}  
Extragalactic rotation measures have been used with great success to probe the magnetic field of the Milky Way in the Galactic plane \citep{brown07}, as well as the Large and Small Magellanic Clouds \citep{gaensler05,mao08}.  This is the first study of RMs as a probe of HVCs.

Here we present an analysis of the RMs towards an HVC in the Magellanic Leading Arm, HVC $287.5+22.5+240$, which shows a possible signature of magnetic fields associated with the HVC.  We estimate the coherent and random magnetic field strengths implied for this HVC and discuss the implications for its stability.  

%----------------------------------------------------------
\section{Data}
\label{sec:data}
%----------------------------------------------------------
We compare atomic hydrogen (\HI) data from the Leiden-Argentine-Bonn (LAB) survey \citep{kalberla05} and the Galactic All-Sky Survey \citep[GASS;][]{mcgriff09,kalberla10} with rotation measures derived  by \citet{taylor09} from the  NRAO Very Large Array Sky Survey \citep[NVSS;][]{condon98}.  LAB covers the entire sky with an angular resolution of $36\arcmin$, a spectral resolution of $1.3$ \kms\ over the range $-400 \leq v \leq +450$ \kms\ with a sensitivity of 70-90 mK.  GASS is a new $21$ cm \HI\ survey of the sky with $\delta \leq +1\arcdeg$, fully corrected for stray-radiation, with an angular resolution of $16\arcmin$, a spectral resolution of $1$ \kms\ and a sensitivity of $\sim 50$ mK.  We used GASS for searches of $\delta < 0\arcdeg$ and in the analysis of HVC $287.5+22.5+240$  presented here.  

The \citet{taylor09} rotation measure catalog contains more than 37,000 sources with $\delta >-40\arcdeg$.  The RMs were calculated from the two NVSS 42 MHz wide frequency bands at 1364.9 MHz and 1435.1 MHz.  \citet{taylor09} estimate that the typical error on the RMs is  $\sim 2~{\rm rad~m^{-2}}$, however the errors on individual RMs can be larger and with some data (mainly at low Galactic latitudes) also subject to large RM ambiguities.  

%----------------------------------------------------------
\section{Results}
\label{sec:results}
\begin{figure}
\centering
\includegraphics[angle=-90,width=6in]{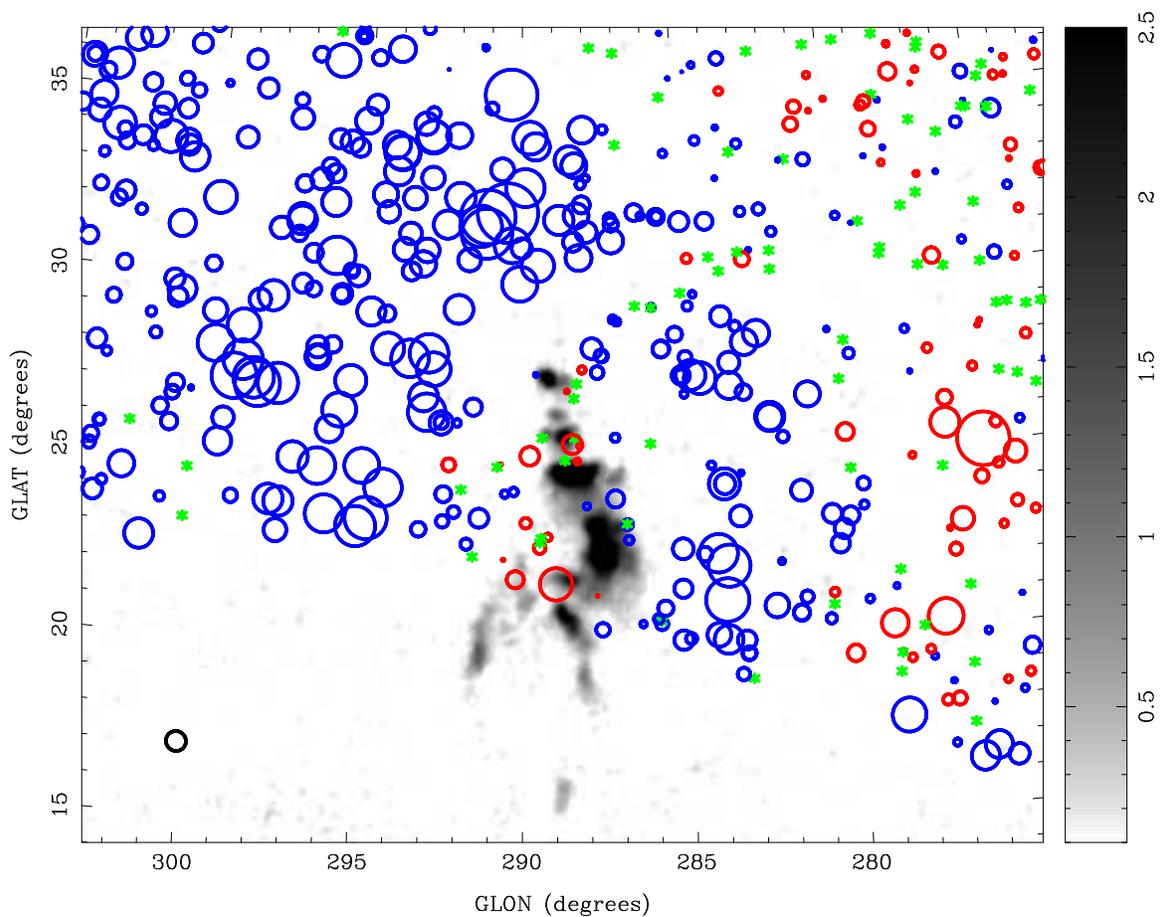}
\caption[]{\HI\ emission at $v_{LSR}=+238$ \kms\ of the region surrounding  HVC $287.5+22.5+240$ overlaid with \citep{taylor09} RMs.
  The \HI\ emission colour scale ranges from $0.1 - 2.5$ K, as shown in
  the wedge at the right.  Positive RMs are plotted in red, negative RMs are plotted in blue.   Values consistent with zero are shown with green stars.  The circle diameter is proportional to the magnitude of the rotation measure; the black circle at the bottom left of the image shows an RM of $50~{\rm rad~m^{-2}}$. 
\label{fig:hvc+rm_big}}
\end{figure}

%----------------------------------------------------------
We searched for morphological agreement between the sign and magnitude of \citet{taylor09} RMs and \HI\ images of all major HVC complexes \citep{wakker91} and 10 large  ($\gtrsim 15~{\rm deg^2}$), isolated HVCs.  The 10 isolated HVCs were selected at latitudes $|b| > 15\arcdeg$ to avoid strong polarization and depolarization signatures from the Galactic plane, which cause ambiguities in interpreting the rotation measurements.  We found a morphological agreement between the \HI\ and RM towards three HVCs or complexes.  We searched complementary all-sky surveys in the optical, far IR, and radio for the presence of foreground structures.  These maps included: radio continuum at 408 MHz \citep{haslam81,haslam82},  IRAS 100 ${\rm \mu m}$ \citep{wheelock94} and \ha\ from SHASSA \citep{gaustad01} and  WHAM \citep{haffner03}, where the \ha\ emission is likely dominated by structures within $\sim 1$ kpc of the Sun. Towards the HVC Complex M we found an agreement, which we attributed the to the well-known foreground Galactic radio loop III \citep{berkhuijsen71}.  Another RM agreement exists towards one of the HVCs in the Complex GCP near ($l,b,v$) = ($41^{\circ}, -22^{\circ},+100~{\rm km~s^{-1}}$), but this may be attributed to \ha\ emission in the solar neighborhood ($v \approx 0 {\rm km~s^{-1}}$).  Finally, towards one HVC in the Magellanic Leading Arm complex LA II, HVC $287.5+22.5+240$ we found evidence for agreement between the RMs and the \HI\ distributions (see Figure \ref{fig:hvc+rm_big}) with no obvious confusing foreground object in radio continuum \ha\ or infrared emission.  We also found no other foreground \HI\ structure in the GASS data that matched the morphology of these RMs.  

HVC $287.5+22.5+240$ is kinematically associated with the Leading Arm, showing a clear velocity connection to the rest of the Magellanic system \citep[e.g.][]{bruens05}.  This HVC has a classic head-tail structure, suggesting that it is moving towards higher latitudes.  The cloud has a mean \HI\ column density of a few $\times 10^{19}~{\rm cm^{-2}}$ and a peak column density of $2.7 \times 10^{20}~{\rm cm^{-2}}$.  The distance to the HVC is not known.  However simulations of the Magellanic system \citep{yoshizawa03,connors06} place the Leading Arm closer than the Magellanic Clouds or the Magellanic Stream, which are assumed to be at distances of 50 - 60 kpc.  \citet{mcgriff08} estimated a kinematic distance of 20 kpc for another Leading Arm HVC at $b\approx 0\arcdeg$.  In the analysis below we assume a distance of $d\sim 30$ kpc.  At this distance the HVC has plane-of-sky dimensions of $1~{\rm kpc} \times 5~{\rm kpc}$ and a total neutral mass of $\sim 8000~{\rm M_{\odot}}$.

\begin{figure}
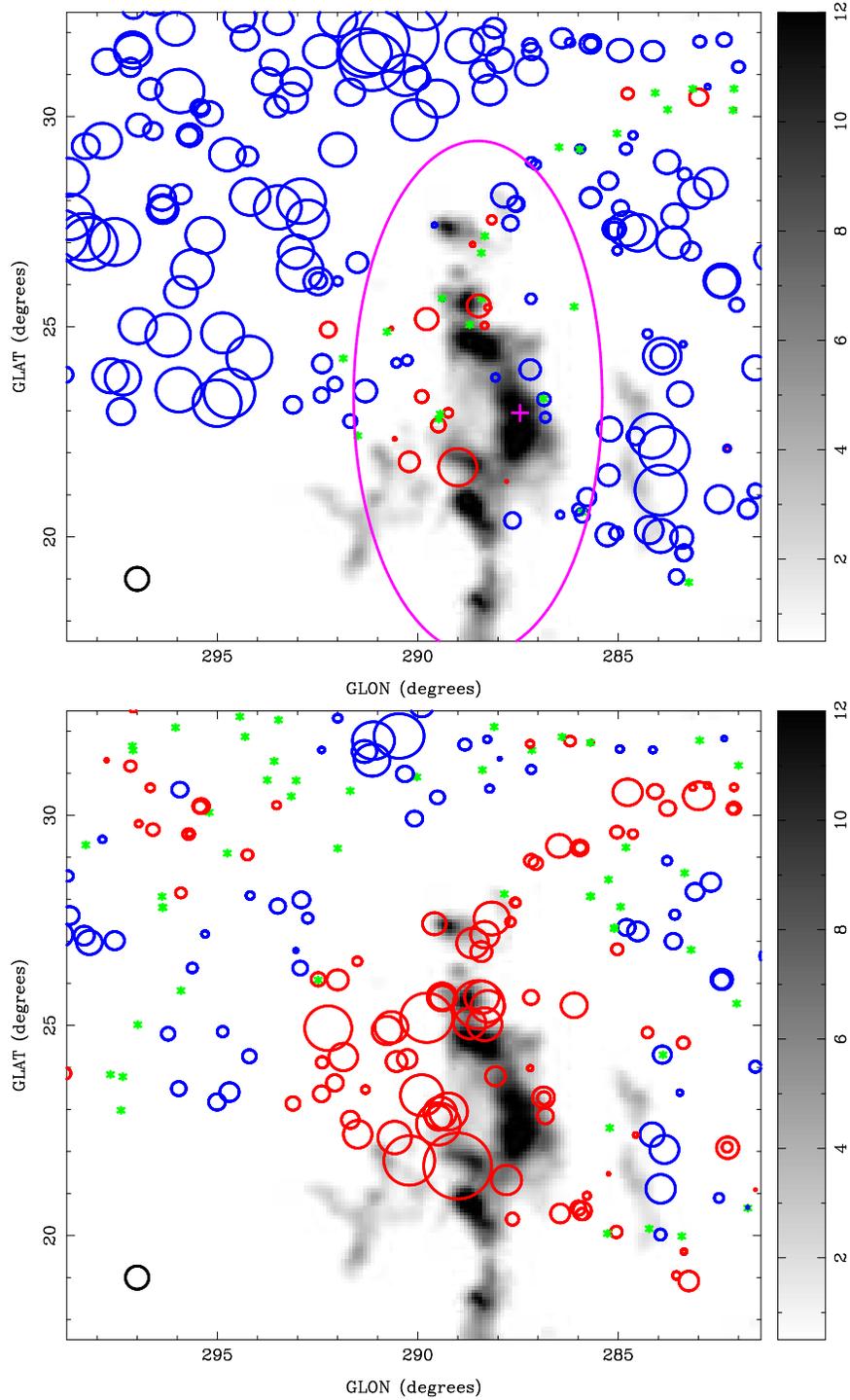

\centering
\includegraphics[angle=-90,width=4.5in]{f2a.ps}\\
\includegraphics[angle=-90,width=4.5in]{f2b.ps}
\caption[]{\HI\ column density of   HVC $287.5+22.5+240$ overlaid with \citep{taylor09}  RMs.
  The \HI\ emission grey scale ranges from $0.5 - 12$ in units of $10^{19}~{\rm cm^{-2}}$, as shown in the wedge at the right.  The symbols are the same as in Figure \ref{fig:hvc+rm_big} .  The purple cross in the top panel marks the location of the QSO NGC 3783 and the ellipse marks the area used to define on-source as described in \S \ref{sec:results}.   The top panel shows the measured RMs and the bottom panel shows the same RMs from which a quadratic surface fit to the off-source RMs has been subtracted.  
  \label{fig:hvc+rm}}
\end{figure}

Figure \ref{fig:hvc+rm_big} shows the \HI\ emissivity of HVC $287.5+22.5+240$ near its central velocity, overlaid with the \citet{taylor09} RMs.  The lower third of the image is below the declination limit of the NVSS.  The RMs in the area surrounding the HVC are mostly negative but the RMs coincident with and immediately surrounding the HVC \HI\ emission are of noticeably smaller magnitude.  Figure~\ref{fig:hvc+rm} shows the \HI\ column density in a smaller region, allowing us to examine individual RMs more closely.  We define an ``on-source'' region to be an ellipse of $6.2\arcdeg \times 12.0\arcdeg$ centred at ($l,b=288.5\arcdeg, 23.3\arcdeg$), which covers an area extending to $1.8\arcdeg$ from the $N_H=2.8\times 10^{18}~{\rm cm^{-2}}$ contour to account for a potential extended ionized halo around the \HI\ as observed in other HVCs \citep[e.g.][]{fox10}.  We define an ``off-source'' region as an elliptical annulus centered on the on-source ellipse, but with twice the area of the on-source ellipse (see Figure \ref{fig:hvc+rm}).  There are 48 RM measurements in the on-source region with 29 directly overlapping the region where $N_H>2.8\times 10^{18}~{\rm cm^{-2}}$.  The on-source RMs have a median and rms ($\sigma$) of $-8.3\pm28.8~{\rm rad~m^{-2}}$, whereas the off-source values are $-48.9\pm36.2~{\rm rad~m^{-2}}$, giving a difference between the two of 1.4$\sigma$.  A K-S test finds that the two distributions are different at the 99\% confidence level.  As seen in Figure~\ref{fig:hvc+rm}, the RMs are predominantly negative through the right half of the HVC and positive through the left half.  The on-source gradient from $\sim +50 ~{\rm rad~m^{-2}}$ on the left side to $\sim -50 ~{\rm rad~m^{-2}}$ on the right side is opposite to the gradient seen in the surrounding RMs.

To assess the likelihood that the association between the RMs and the HVC is real we have used the \citet{taylor09} all-sky RM database to estimate the probability of finding an enhancement in RMs on the same angular scale as the HVC.  We placed 10,000 ellipses with the same areas described above at random locations in the Northern sky and with random position angles.  We found that only 1\% of the ellipses enclosed areas exhibiting an RM enhancement of greater than $ 1.0\sigma$.  Even fewer than 1\% show an enhancement as large or larger than the $1.4\sigma$ observed on HVC $287.5+22.5+240$.  The fact that RM enhancements on this angular scale are rare, coupled with finding only 1 out of 27 HVCs to have a distinct RM-\HI\ morphological agreement, suggests that the two are causally related. More data are needed to confirm this association (see \S\ref{sec:conclusions}).

The observed RM may contain components from the intergalactic background, the HVC and the Milky Way foreground.  To remove the contribution from the Milky Way foreground and the intergalactic medium we fit a quadratic surface to the 304 sources in the area $282\arcdeg \leq l \leq 310\arcdeg$, $b\leq +35\arcdeg$ outside the boundary of the HVC defined above and subtract the fit from the on-source RMs. The fit is limited to $l>282\arcdeg$ to avoid the strong positive RM feature at $l\sim 278\arcdeg$, which is correlated with positions of enhanced foreground H$\alpha$ emission in SHASSA \citep{gaustad01}.  The residual RMs (lower panel of Figure~\ref{fig:hvc+rm}) show a clear excess of RM on, and immediately surrounding, the HVC.  The residual on-source RMs show a gradient from $\sim +44~{\rm rad~m^{-2}}$ at $l=287\arcdeg$ to $\sim +75~{\rm rad~m^{-2}}$ at $l=291\arcdeg$, with a mean value of $+55\pm27~{\rm rad~m^{-2}}$.  Different estimations of the off-source RMs, including a bi-linear fit and a simple median, give mean residual on-source RMs in the range $+50$ to $+70~{\rm rad~m^{-2}}$.

\subsection{Magnetic Field Strength}
\label{subsec:bfield}
If we assume that the residual RM structure is due to HVC $287.5+22.5+240$ then we can use the foreground-subtracted rotation measures and an estimate of the electron density in the cloud, $n_e(l)$, to estimate the average coherent magnetic field strength in the HVC.  The full form of $n_e(l)$ is not known and we work instead with the product of an average electron density, $n_0$, and the assumed path length, $L$, through the HVC.

\subsubsection{Electron Column Density Estimates}
For HVC 287.5+22.5+240 we have used two methods to estimate the electron density: observations of the Balmer-$\alpha$ (\ha) recombination line of hydrogen and absorption metal line measurements in the UV and Far-UV towards a background quasar.  The detection of \ha\ emission from HVCs requires a combination of high spectral resolution and very high surface brightness sensitivity \citep[see][]{tufte02, putman03}.  We used the Wisconsin H-Alpha Mapper South \citep[WHAM-South][]{haffner10}, a Fabry-Per{\'o}t facility now located at the Cerro Tololo Interamerican Observatory. WHAM-South records the average spectrum over a 1$^\circ$ circular field within a 200 \kms\ wide window at a spectral resolution of 12 \kms\ \citep{reynolds98}.

A set of observations were taken for five directions towards HVC 287.5 + 22.5 + 240; they are spread throughout the cloud and biased toward high \HI\ column densities (N$_{\rm{HI}} > 2 \times 10^{19}~$cm$^{-2}$).  In order to remove the contribution from telluric emission lines, a concurrent set of observations were taken few degrees away from each target; these directions show no \HI\ emission at the velocity of the HVC.  An average of the `off' field spectra were subtracted from the target spectra.  The data were reduced using a standard WHAM data pipeline \citep[see][]{madsen06}. The surface brightness and velocity were calibrated using synchronous observations of the bright HII region surround $\lambda$ Ori, and linked to the absolutely calibrated observations from the WHAM Northern Sky Survey \citep{haffner03}. The data have an uncertainty of 10\% and 2 \kms\ in surface brightness and velocity, respectively.

\begin{figure}
\centering
\includegraphics[width=6in]{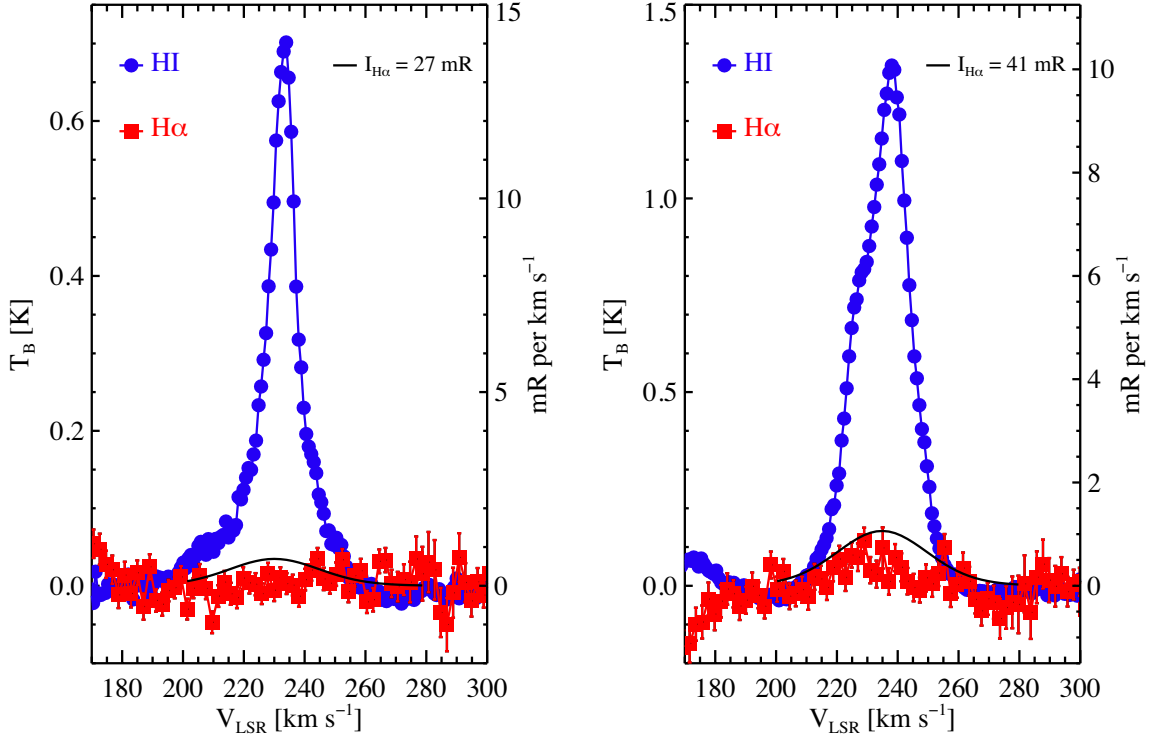}
\caption[]{\HI\ and \ha\ spectra  toward the HVC $287.5+22.5+240$.  The left
and right panels show spectra centred at Galactic coordinates
$(288.90^\circ, +26.95^\circ)$ and $(288.75^\circ, +24.65^\circ)$,
respectively.  The \ha\ spectra are shown in units of milli-Rayleigh per \kms; the \HI\ spectra are shown in units of Kelvin.   The \HI\ spectra (blue circles) are from the GASS survey;
the \ha\ spectra (red squares) are new data from WHAM-South.  A 3$\sigma$ upper 
limit to the strength of an undetected \ha\ line in each panel is shown as 
a dark solid line, with the peak \ha\ strength given in the upper right of each panel. 
\label{fig:wham}}
\end{figure}

A sample of results from our velocity-resolved WHAM-South observations are shown in Figure \ref{fig:wham}.  The \ha\ spectra are shown as red squares, with \HI\ spectra from GASS overlaid as blue circles.  The velocity is given in the reference frame of the kinematic local standard of rest.  The \ha\ spectra are shown in units of milli-Rayleigh\footnotemark \footnotetext{1 Rayleigh = $10^6$/4$\pi$ photons s$^{-1}$ cm$^{-2}$ sr$^{-1}$ = 5.7 x 10$^{-18}$ ergs cm$^{-2}$ s$^{-1}$ arcsec$^{-2}$ at \ha} per \kms; the \HI\ spectra are shown in units of brightness temperature.  A 3$\sigma$ upper limit to the strength of an (undetected) \ha\ line is given in the legend and shown as a dark solid line. The upper limits were calculated by fixing the line center at the velocity centroid of the \HI\ data and by assuming a line width (FWHM) of 30 \kms, typical of \ha\ detections in other HVCs (see \citet{gallagher03} for a details of the method).  The two panels in Figure \ref{fig:wham} represent the minimum and maximum upper limits to the \ha\ line for all five targets; the average upper limit is $0.034$~mR.  An additional correction for extinction by interstellar dust must be applied. The dust maps of \citet{SFD98} show the reddening toward the cloud is $E(B-V) = 0.10$~mag, implying an upward correction to the \ha\ flux of $< 27$\% \citep{finkbeiner03}.  We place an upper limit on the \ha\ surface brightness of the HVC of $I_{H\alpha}\leq 0.04$ Rayleighs.  

The low H$\alpha$ flux is not surprising as the HVC is quite distant from the Milky Way and at a relatively low Galactic latitude; both factors may reduce the ionizing flux incident on the cloud \citep{bland-hawthorn99}.  Using the relation $EM = 2.75 (T/10^4~{\rm K})^{0.9} I_{H\alpha} = n_0^2 f L$, where $f$ is the volume filling factor of the gas with characteristic density $n_0$  \citep{reynolds91}, we can set a limit on the electron density.  Assuming constant occupation length, $fL$, the free-electron column density or dispersion measure, $DM=n_0 f L$, is related to the emission measure as $\langle DM \rangle = (f L \, \langle EM \rangle)^{0.5}$.  The total path length through the HVC is unknown, but if we assume the depth along the line of sight is comparable to the plane-of-sky width of the RM patch, then $L \sim 2.5$ kpc at an assumed distance of 30 kpc.  Given the WHAM H$\alpha$ limit we estimate $DM \lesssim 16.5 f^{0.5} ~{\rm pc~cm^{-3}}$ or a column density of $N_{\rm HII} \lesssim 3.6 \times 10^{19}~{\rm cm^{-2}}$ for $f=0.5$.

We also estimate $N_{\rm HII}$ from measurements of the UV and Far-UV absorption metal lines through the HVC towards a background QSO.  Archival HST and FUSE absorption line measurements towards NGC 3783 at $(l,b)=(287.45\arcdeg,+22.95\arcdeg$) show \ion{Si}{2}, \ion{Si}{3}, \ion{Si}{4}, \ion{Fe}{2}, \ion{S}{2} and \ion{C}{4} absorption at the velocity of HVC $287.5+22.5+240$ \citep{wakker01,shull09}.  To estimate $N_{\rm HII}$ we have constructed a photo-ionization model using version C08.00 of Cloudy \citep{ferland98}.  The model assumes a plane-parallel cloud with an \HI\ column density of $N_{\rm HI} = 1.2 \times 10^{20}~{\rm cm^{-2}}$ (as measured from GASS), a metallicity of $0.25$ times the standard solar gas phase abundances \citep{lu98} and standard extragalactic and Milky Way radiation fields \citep{fox05}.  From the absorption lines available our simple model gives an estimate of $N_{\rm HII} = 1 - 4 \times 10^{19}~{\rm cm^{-2}}$, consistent with the H$\alpha$ limit.  This implies an ionization fraction within the HVC of $N_{\rm HII} / N_{\rm HI+ HII} = 0.08 - 0.25$, which is consistent with measurements of the ionization fraction in other HVCs \citep{lehner04}.

\subsubsection{Coherent and Random components of $B$}
Given the estimate $N_{\rm HII}\lesssim 3.6 \times 10^{19}~{\rm cm^{-2}}$, Equation \ref{eq:rm} can be re-written as $\langle B_{\parallel} \rangle= 3.8\times 10^{18} \langle RM_{HVC} \rangle/N_{\rm HII}$, where $\langle RM_{HVC}\rangle = 55~{\rm rad~m^{-2}}$ is the average rotation measure attributable to the HVC.  The implied coherent magnetic field strength is $\langle B_{\parallel}\rangle \gtrsim +6~{\rm \mu G}$ with the field pointed towards the observer.  A randomly oriented magnetic field of strength $B_r$ will induce fluctuations in RM that are reflected in the dispersion of the measured RMs, $\sigma_{RM} = 27~{\rm rad~m^{-2}}$.  We can estimate an upper limit to $B_r$ by assuming that all of $\sigma_{RM}$ is due to a random magnetic field.  This limit neglects variations in $N_{\rm HII}$, residual variations in the foreground RM and the variance in the intrinsic RMs of the extragalactic background sources, all of which will serve to reduce any derived value of $B_r$.  We find $B_r = \sqrt{3} \sigma_{RM}3.8\times 10^{18} / N_{\rm HII} \leq 5~{\rm \mu G}$, where the $\sqrt{3}$ term comes from the definition of a normal distribution of $N$ measurements about a mean of zero with a variance of $N/3$.  This implies $B_{r}/B_{\parallel} < 0.8$ and suggests that the contribution of the random field to the total field is subdominant to the coherent field, unlike the LMC \citep{gaensler05} and SMC \citep{mao08}.

\section{Discussion}
\label{sec:discussion}
Several authors discuss how HVCs are able to survive the passage through the Milky Way halo to encounter the disk \citep[e.g.][]{heitsch09,quilis01}.  Observationally it is clear that many HVCs are disturbed by the ram pressure interaction with the halo, producing head-tail structures \citep[e.g.][]{bruens00,peek07}.  Simple calculations show that for an HVC moving supersonically through the halo with number  density, $n_h$, the typical ram pressure destruction timescale for a cloud of radius, $r_c$, density, $n_c$ and travelling at a velocity $v_c$ is 
\begin{equation}
t_{bc} = \sqrt{\frac{n_c}{n_h}} \frac{2 r_c}{v_c}
\end{equation}
\citep{jones96}.  For HVC 287.5+22.5+240 we only have a measurement of the radial component of $v_c$.  However, we can assume that the Leading Arm has a space velocity similar to the velocity of the Magellanic Clouds themselves, i.e.\ $v_{LMC}\approx 370$ \kms\  \citep{kallivayalil06a}.  The density of the ambient medium, $n_h$, is poorly constrained, we assume $n_h\sim 10^{-4}~{\rm cm^{-3}}$ \citep{bregman09}.  For $r_c\sim 1$ kpc and $v_c\sim 300$ \kms, we estimate that the destruction time is $t_{bc}\sim 25$ Myr, much less than the $\sim 500 - 1000$ Myr estimated for the Leading Arm material to have moved away from the Magellanic Clouds \citep{connors06}.  Even if the ambient density is a factor of ten smaller, the destruction time is still much less than the HVC travel time.  

Surface tension from magnetic fields may be a means of stabilizing an HVC \citep{konz02}.  To estimate the magnetic field strength required to stabilize the cloud at its apex we can balance the ram pressure with magnetic pressure:
\begin{equation}
1/2 \, m_H n_h v_c^2 = \frac{B^2}{8\pi}.
\end{equation}
From this we estimate that a magnetic field of $B\sim 4~{\rm \mu G}$ would be sufficient to balance the ram pressure for HVC 287.5+22.5+240.  The coherent line-of-sight magnetic field of $\gtrsim 6~{\rm \mu G}$ proposed in \S \ref{subsec:bfield} is therefore consistent with the field required to dynamically stabilize the cloud against ram pressure.

A magnetic field can also provide a thermal barrier between the cold HVC and the hot Galactic halo and hence increase its lifetime.  A 1 kpc cold \HI\ cloud bathed in a $10^{5-6}$ K halo will suffer from an evaporation mass loss rate of $6.5 - 20\times 10^{-6}~{\rm M_{\odot}~yr^{-1}}$ \citep[]{mckee77b}, which implies a cloud lifetime of $300 - 1000$ Myr for HVC 287.5+22.5+240.  In practice this is likely an under-estimate because a moving cloud will suffer thermal ablation \citep{bland-hawthorn07}.  Given the time since separation from the Magellanic Clouds, our estimate of the cloud's current mass would be about half its original mass. A magnetic barrier wrapped around the head of the moving cloud reduces thermal conduction almost completely, but depends strongly on the field geometry \citep{balbus86}, which is unknown.

HVCs are also vulnerable to instability processes, particularly the Kelvin-Helmholtz and the Rayleigh-Taylor instability \citep{jones94}.  Analytically, a magnetic field parallel to the direction of motion can stabilize an HVC against these effects \citep{chandrasekhar61}; two-dimensional MHD simulations of HVCs support the analytic results \citep{konz02}.  Three-dimensional MHD simulations of  the physically similar ``magnetic draping'' in cluster environments \citep{dursi08} show that the Kelvin-Helmoholtz and Rayleigh-Taylor instability are indeed supressed in the presence of a parallel magnetic field, but remain in the direction transverse to the magnetic field and that in this direction the Kelvin-Helmholtz instability dominates in the destruction of the projectile.  The orientation of the halo magnetic field near HVC  287.5+22.5+240 is unknown.  In some external galaxies, the lower halo field extends from the disk field and is roughly parallel to the disk \citep[e.g.][]{braun10} and perpendicular to the direction of motion.

It is not clear how a strong coherent magnetic field is established in an HVC.  Most authors assume that the Leading Arm gas comes from either the Large or Small Magellanic Cloud (SMC).  Both the LMC and SMC are estimated to have weak magnetic fields that are coherent on scales greater than $\sim 100$ pc with strengths of $\sim 1.2~{\rm \mu G}$ and $0.2~{\rm \mu G}$, respectively \citep{gaensler05,mao08}. The HVC may have possessed a weak ordered magnetic field when it was stripped off the clouds, and then would have required amplification.  The \citet{konz02} 2-D magneto-hydrodynamic simulations have shown that even an unmagnetized HVC travelling through a weakly magnetized ($\sim 0.3~{\rm \mu G}$) halo can develop a magnetized barrier layer, whose strength is on the order of ten times that of the halo field.  They find that the resultant $3~{\rm \mu G}$ field is more than sufficient to provide thermal and dynamical stabilization for the HVC.  Observational estimates of the magnetic field strength in the halo are few and limited to several kiloparsecs of the Sun.  \citet{sun08} estimate halo field strengths parallel to the disk of $0.1 - 1~{\rm \mu G}$ over the range $0 \leq z \leq 3$ kpc at a Galactic radius of $\sim 20$ kpc, which is where the Leading Arm appears to cross the disk \citep{mcgriff07}.  Although the halo field is likely much less than this at the current position of the HVC, if the HVC traversed the Galactic disk and we assume a magnetic field amplification of a factor of ten \citep{jones96,konz02}, then we might expect to find a magnetic field of magnitude $1-10 {\rm \mu G}$ in HVC 287.5+22.5+240.

%----------------------------------------------------------
\section{Conclusions}
\label{sec:conclusions}
%----------------------------------------------------------
We have compared extragalactic rotation measures with the \HI\ morphology of a large HVC associated with the Magellanic Leading Arm, HVC 287.5+22.5+240.  We find that the RMs on-source are significantly different from the surrounding RMs and estimate the probability that the RM/HVC association is real at $>99$\%.  We suggest that the RMs are indicative of a coherent magnetic field in the HVC.  Using the RMs, corrected for a foreground Milky Way contribution, and an H$\alpha$ emission upper limit, together with QSO absorption lines to estimate the electron content of the HVC, we have estimated a coherent line-of-sight magnetic field strength in the HVC of $\gtrsim 6~{\rm \mu G}$.  The HVC is a promising target for HI Zeeman observations to give an accurate and, more importantly, {\em in situ} measurement of the magnetic field of an HVC.   If this detection is confirmed, a magnetic field of this magnitude has important implications for HVC stability and ultimately for the accretion of mass in the Milky Way.  

Of the 27 HVCs and HVC complexes searched, HVC 287.5+22.5+240 is the only one for which an RM enhancement was found that could not be clearly associated to a foreground object.  It is not clear why HVC 287.5+22.5+240  is detectable when other HVCs were not.  \ha\ and radio continuum foregrounds are favourable for HVC 287.5+22.5+240 because there is no strong, highly structured emission in the foreground to mask an HVC RM signature.  One reason for detection may in fact be the Magellanic origin of this HVC.  If the cloud originated from magnetised material and possesses a large space velocity, as indicated by head-tail structure and models of the Magellanic system, then we might speculate that the magnetic field strength is due to a combination of compression of the HVCs initial field and compression of a weak Milky Way halo field, through which the HVC is moving.   If this speculation is correct then the rest of the Magellanic Stream, most of which is south of the \citet{taylor09} database, would be an ideal target for future RM searches of HVCs.  Additional searches for magnetic fields in Southern HVCs are indeed planned with the hundreds of thousands of RMs that will be derived from an Australian SKA Pathfinder \citep[ASKAP][]{johnston07} all-sky survey of polarized point sources, POSSUM \citep{gaensler10}.  

\acknowledgements We are grateful to Tobias Westmeier for providing a FITS version of his LAB High-Velocity Sky map, to Marijke Haverkorn for useful conversations and Andy Fox for advice on CLOUDY models.  The Wisconsin H-Alpha Mapper South is funded by the National Science Foundation through award AST-0607512.  WHAM-South is the result of the dedicated efforts of a team of people including L. Matt Haffner, GJM, Alex Hill, Kurt Jaehnig, Ed Mierkiwiecz, and Ron Reynolds {\it Facilities:} \facility{Parkes}, \facility{WHAM}, \facility{VLA}

%-------------------------------------------- 
%Bibliography
%--------------------------------------------
\bibliographystyle{apj} 
%\bibliography{references} %~mcg/tex/references.bib, bibtex file 

\normalsize

%---------------------------------------------
% Figures and Tables
%---------------------------------------------

\clearpage

\end{document}